\documentclass{webofc}
\usepackage{subfig}
\usepackage{multirow}
\usepackage[varg]{txfonts}   
\begin{document}
\title{On the generation of random ensembles of qubits and qutrits}
%
%
\subtitle{Computing separability probabilities 
for fixed rank states}

\author{\firstname{Arsen} \lastname{Khvedelidze}\inst{1,2,3,4}\fnsep\thanks{\email{akhved@jinr.ru}} \and
        \firstname{\textit{Ilya}} \lastname{Rogojin}\inst{4}\fnsep\thanks{\email{virus-atl@inbox.ru}} 
}

\institute{A.Razmadze Mathematical Institute, 
           Iv.Javakhishvili Tbilisi State University,                Tbilisi, Georgia
\and
           Institute of Quantum Physics and 
           Engineering Technologies, 
           Georgian Technical University, Tbilisi, Georgia
\and
           National Research Nuclear University,
           MEPhI (Moscow Engineering Physics Institute),
           Moscow, Russia
\and
           Laboratory of Information Technologies,  
           Joint Institute for Nuclear Research,
           Dubna, Russia
          }

\abstract{The question of the generation of random mixed states is discussed, aiming for the computation of probabilistic characteristics of  composite finite dimensional quantum systems. Particularly, we consider the generation of the random Hilbert-Schmidt and Bures ensembles of qubit and qutrit pairs and compute the  corresponding probabilities to find a separable  state among the states of a fixed rank.}

\maketitle

\section{Introduction}
\label{intro}

The presentation addresses a special topic of quantum theory of finite dimensional systems that is of prime importance in the theory of quantum information and quantum communications. Namely, we study the problem of calculation of the probability for a random state of a binary composite quantum system, such as qubit-qubit or qubit-qutrit pairs, to be  a separable or an entangled state 
\footnote{For definitions and references we quote the  comprehensive review \cite{BengtssonZyczkowskiBOOK} and a recent paper by P.Slater \cite{Slater}.}. 
According to the “Geometric Probability Theory” \cite{KlainRota1997}, this «separability/entanglement probability», is determined by  the relative volume of separable states with respect to the  volume of the whole state space. 
In order to avoid subtle numerical calculation of the multidimensional integrals, (for a generic 2-qubit case the integrals are over the semi-algebraic domain of 15-dimensional Euclidean space), the Monte-Carlo ideology with a specific method of generation of random variables has been used (see e.g.. \cite{Braunstein1996}-\cite{KhvedRogoj2015},   and references therein). Our report aims to present several results on the numeric studies of separability probabilities for random qubit-qubit  and qubit-qutrit  Hilbert-Schmidt and Bures ensembles. Particularly, the distribution properties of the “probability of entanglement” will be described with respect to the subsystem's Bloch vectors for the qubit-qubit and qubit-qutrit  random  Hilbert-Schmidt and Bures states of all possible ranks.  

\section{The separability probability}

There are two ways to analyze the probabilistic aspects of entanglement characteristics, such as the so-called \textit{separability probability}, $\mathcal{P}_{\mathrm{sep}}\,.$ The latter is the probability to find a separable state among all possible states. The first approach is to   consider the state space of a quantum system as the Riemannian space and, following the strategy of the theory of geometric probability, identify the separability probability with with a two volumes ratio:
\begin{equation}\label{eq:SepProbVolumes}
\mathcal{P}_{\mathrm{sep}}= \frac{\mathrm{Vol}\left(\mathrm{Separable~states}
\right)}{\mathrm{Vol}\left(\mathrm{All~states}\right)}\,.
\end{equation}
The second approach consists of dealing with the ensemble of random states, whose distribution is in correspondence with the volume measure in (\ref{eq:SepProbVolumes}). In this case the separability probability can be equivalently written as:   
\begin{equation}\label{eq:SepProbNumber}
\mathcal{P}_{\mathrm{sep}}= \frac{\mathrm{Vol}\left(
\mathrm{Number~of~separable~states~in~ensemble}
\right)}{\mathrm{Vol}\left(\mathrm{Total~number~of~states~
in~ensemble}\right)}\,.
\end{equation}
The first method is suitable for analytic calculations, however it presents a big computational issue (see e.g., \cite{Slater} and references therein).
The second method allows us to use highly effective modern numerical computational techniques. Below we will give results on the separability probability of the qubit-qubit and qubit-qutrit systems obtained within the latter  approach. 

\section{Generating the Hilbert-Schmidt and Bures random states}

Having in mind calculations of the separability probability of states with different ranks we shortly present the algorithm for the generation of density matrices from the Hilbert-Schmidt and the Bures ensembles of a $n\--$level quantum system. 
\begin{itemize}
\item  The first step in the construction of both ensembles is the generation of complex $n\times n$ matrices $\mathrm{Z}$ from the Ginibre ensemble, i.e., the matrices whose
entries have real and imaginary parts distributed as {normal random variables};
\item Density matrices $\varrho_{\mathrm{HS}} $ describing states from the Hilbert-Schmidt 
ensemble are: 
\begin{equation}\label{eq:HSGen}
\varrho_{\mathrm{HS}} = \frac{\mathrm{Z}\mathrm{Z}^+}{\mathrm{tr}\left(\,\mathrm{Z}\mathrm{Z}^+\right)}\,;
\end{equation} 
\item Density matrices $\varrho_{\mathrm{B}}$ from the Bures ensemble are generated with the aid of the Ginibre matrices $Z$ and matrices $U\in SU(n)\,, $ distributed over the $SU(n)$ group according to the Haar measure,
\begin{equation}\label{eq:BuresGen}
\varrho_{\mathrm{B}} = \frac{\left(I+U\right)\mathrm{Z}\mathrm{Z}^+\left(I+U^+\right)}{\mathrm{tr}\left(\,\left(I+U\right)\mathrm{Z}\mathrm{Z}^+\left(I+U^+\right)\right)}\,. 
\end{equation}
\end{itemize}
Since we are interested in studying the entanglement of states with a fixed rank of density matrices, the above algorithm requires a certain specification. For states of a lower rank than the maximal one we proceed as follows.

\noindent $\bullet${\bf Rank-3 states}$\bullet$
We start with the generation of complex, rank-3 Ginibre matrices. It is known that any such matrix admits the following representation: 
\begin{equation}\label{eq:rank3}
Z= P_Z
\left(\begin{array}{@{}c|c@{}}
 {\large A_{}} &
  \begin{matrix}
  x_{1} \\
  x_2 \\
  x_3 
  \end{matrix}
\\ \hline
  \begin{matrix}
  y_1 & y_2 & y_3
  \end{matrix}
  &  D
\end{array}\right)Q_Z\,,
\end{equation}
In (\ref{eq:rank3}) permutations of entries by $P_Z$ and $Q_Z$ are such that $A$ is a regular element of the Ginibre ensemble of $3\times3$ matrices. Apart from that, if 3-tuples $ Y=(y_1, y_2, y_3) $ and $ X=(x_1, x_2, x_3)$ are composed of the normal random complex variables, while  the entry $D$ is   \[D = Y\, A^{-1} X\,,\] 
we arrive at a rank-3 random matrix $Z$ from the Ginibre ensemble of complex $4\times 4$ matrices. 

\noindent $\bullet${\bf Rank-2 states}$\bullet$
Similarly, the  $4\times 4$ Ginibre matrix of rank-2  can be written  as, 
\begin{equation}\label{eq:rank2}
Z= P_Z
\left(\begin{array}{@{}c|c@{}}
  A & B
\\ \hline
  C &  D
\end{array}\right)
Q_Z\,,
\end{equation}
where $A, B$ and $C$ are $2\times 2$ complex Ginibre matrices,  
while  $2\times 2$ matrix $D$ reads,  
$D = C\, A^{-1} B\,.$
 
\noindent $\bullet${\bf Rank-1 states}$\bullet$ Finally,  a $4\times 4$ complex Ginibre matrix $Z$ of rank-1 admits the representation: 
\begin{equation}\label{eq:rank1}
Z= P_Z
\left(\begin{array}{@{}c|c@{}}
 {a}  & \begin{matrix}
  y_1 & y_2 & y_3
  \end{matrix}
 \\ \hline 
\begin{matrix}
  x_{1} \\
  x_2 \\
  x_3 
  \end{matrix} & {\large D_{}}
 \end{array}\right)Q_Z\,,
\end{equation}
where $a, x_1, x_2, x_3$ and $y_1,y_2,y_3$ are normal random complex variables, while  $3\times 3$  matrix  $D$ is: 
\[ D = \frac{1}{a}\left(
\begin{array}{ccc}
  x_1y_1 & x_1y_2 & x_1y_3 \\
  x_2y_1 & x_2y_2 & x_2y_3 \\
  x_3y_1 & x_3y_2 & x_3y_3
\end{array}
\right)\,.
\]
Now, using the generic Ginibre matrix and  representations (\ref{eq:rank3})-(\ref{eq:rank1}) for non-maximal rank matrices 
 we can build  either the Hilbert-Schmidt or Bures states of the required rank according to either (\ref{eq:HSGen}) or (\ref{eq:BuresGen}).

\section{Computing the separability probability} 
 
Computation of the sought-for separability probability (\ref{eq:SepProbNumber}) requires the test of generated states on their separability. With this goal the Peres-Horodecki criterion \cite{BengtssonZyczkowskiBOOK} has been used in our calculations. 
 
\begin{table}
\centering
\begin{minipage}{.4\textwidth}
\caption{The separability probability for the maximal rank-4 states  in  qubit-qubit and qubit-qutrit systems}\label{t:1}
\begin{tabular}{|c|c|c|}
\hline
	States & System & Separable    \\
			\hline
            \hline
\multirow{2}{*}{\bf{HS ensemble}} &$2\otimes 2 $ &0.2424 \\         
			&$2\otimes 3$& 0.0270 \\
			\hline\hline
\multirow{2}{*}{\bf{ Bure ensemble}} &$2\otimes 2$ &0.0733\\
		& $ 2\otimes 3 $  &  0.0014  \\
\hline
		\end{tabular}
\end{minipage}\hfill
\begin{minipage}{.4\textwidth}
\caption{The separability probability for 
non-maximal rank qubit-qubit states}\label{t:2}
\begin{tabular}{|c|c|c|}
\hline
	States & Rank & Separable   \\
			\hline
\multirow{3}{*}{\bf{HS ensemble}} & 3 & 0.1652 \\         
			&2& 0 \\
            &1& 0 \\
            \hline\hline
\multirow{3}{*}{\bf{Bure ensemble}}
	& 3 & 0.0494 \\
	&2& 0 \\
    &1& 0 \\
\hline
		\end{tabular}
\end{minipage}\hfill
\end{table}

\subsection{Results of the computations}

First of all, in table~\ref{t:1} we collect results of the calculations of the separability  probability of the generic (rank-4) states from the Hilbert-Schmidt and Bures ensembles of qubit-qubit and qubit-qutrit pairs.
Furthermore, results of the computations of the separability probabilities of non-maximal rank states of a qubit-qubit system, for both ensembles, are given in table~\ref{t:2}.
Here, it is worth making a comment on the zero separability probability for rank-2 and rank-1 states. That can be understood noting the following: 
If a density matrix of 2-qubits $\varrho$ is such, that  $\mathrm{rank}(\varrho) < d_A=\mathrm{rank}(\varrho_A)$,  then  $\varrho$ is not separable 
\cite{RuskaiWener2009}. Here, $\varrho_A$ denotes the reduced density matrix obtained by taking the partial trace with respect to the second qubit. Indeed, since during the generation of rank-2 and rank-1, almost all  reduced matrices are not singular, the above mentioned inequality is true.  
\begin{figure}[htp] 
    \centering
    \caption{The separability probability for the Hilbert-Schmidt qubit-qubit states as a function of the first qubit's Bloch vector.}
    \subfloat[Maximal rank states]{%
\includegraphics[width=0.46\textwidth]{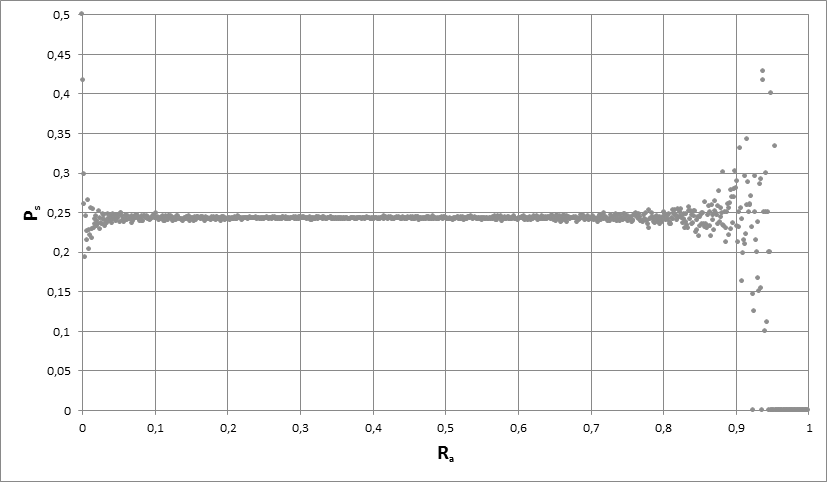}
        \label{fig-1}%
        }%
    \hfill%
    \subfloat[Rank-3 states]{%
\includegraphics[width=0.46\textwidth]{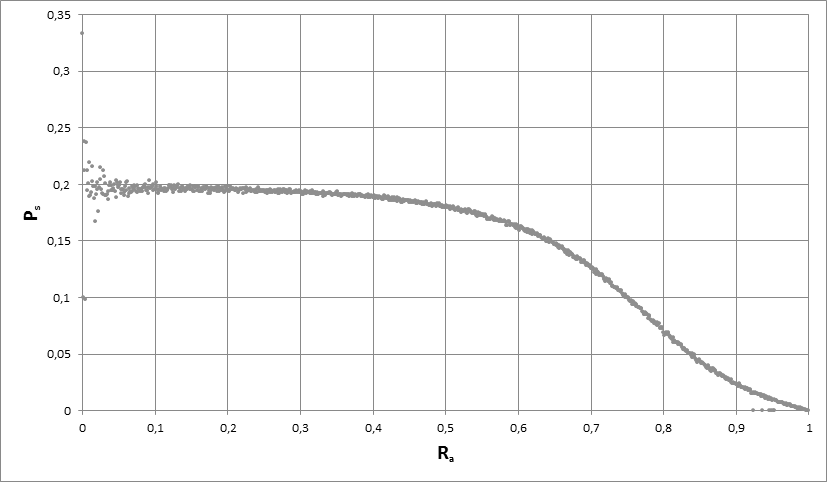}%
        \label{fig-2}%
        }%
\end{figure}

\section{Concluding remarks}

Having the above described algorithms as a tool for the generation of ensembles of random states with different ranks, the diverse characteristics of entanglement in composite quantum systems can be studied. As an example, on figures~\ref{fig-1} and \ref{fig-2}, we illustrate the separability probability in 2-qubit systems as a function of the Bloch radius of the constituent qubit for the states of maximal and sub-maximal ranks. Our analysis shows that the distribution of the separability probability with respect to the Bloch radius of the qubit is uniform for maximal rank states (see figure~\ref{fig-1}), while for the rank-3 states the deviation from the total separability probability is depicted on figure~\ref{fig-2}.


\end{document}